\documentstyle[prl,twocolumn,aps]{revtex}

\begin{document}
\draft

\title{Orientational Ordering in Spatially Disordered Dipolar Systems}

\author{
G. \ Ayton$^*$, M.\ J.\ P.\ Gingras$^\dagger$, and G.\ N.\ Patey$^*$}

\address{
$^*$Department of Chemistry, University of British Columbia, Vancouver,
British Columbia, Canada V6T 1Z1 }

\address{
$^\dagger$TRIUMF, 4004 Wesbrook Mall, Vancouver, British Columbia, Canada V6T
2A3}

\date{\today}

\maketitle

\begin{abstract}

This letter addresses basic questions concerning ferroelectric order in
positionally disordered dipolar materials. Three models distinguished by
dipole vectors which have one, two or three components are studied by computer
simulation. Randomly frozen and dynamically disordered media are considered.
It is shown that ferroelectric order is possible in spatially random systems,
but that its existence is very sensitive to the dipole vector dimensionality
and the motion of the medium. A physical analysis of our results provides
significant insight into the nature of ferroelectric transitions.

\end{abstract}


\narrowtext

Recently, spatially disordered dipolar materials have attracted considerable
attention. Diluted lattices \cite{vug,hochli,xu_log}, fluid phases
\cite{wei,weis} and amorphous frozen ferrofluids \cite{luo} have been examined
experimentally \cite{vug,hochli,luo}, theoretically
\cite{xu_log,cieplak,levin,wpp,dietrich,widom} and with computer simulations
\cite{xu_log,wei,weis}. It has been shown that in the absence of long-range
positional order, the strong spatial-orientational coupling intrinsic to
dipolar forces can lead to interesting new phase behavior. For example, frozen
ferrofluids containing  magnetic particles in a non-magnetic solvent exhibit
magnetic irreversibilities reminiscent of randomly frustrated magnetic spin
glasses \cite{BY}. On the other hand, computer simulations of simple dipolar
fluids clearly indicate the existence of a ferroelectric liquid crystal phase
\cite{wei,weis}.

A simple interpretation of these observations might be as follows. In frozen
ferrofluids, the quenched positional disorder creates random frustration and
the system behaves as a spin glass \cite{BY}. The fluid systems
\cite{wei,weis} differ from the frozen case in that the strongly coupled
translational and rotational degrees of freedom are in full thermal
equilibrium. This allows the development of short-range spatial correlations
resembling those seen in the ferroelectric tetragonal-I lattice \cite{tao}
and, consequently, ferroelectric order develops in the liquid phase
\cite{wei}. In view of these observations, and recalling that perfect crystals
exhibit ferroelectric or antiferroelectric long-range order depending on the
lattice symmetry \cite{tao,lutt}, one might argue that {\em specific} spatial
correlations are required for ferroelectric order.

In recent papers Zhang and Widom \cite{widom} have put forward a mean field
theory for spatially disordered dipolar systems. They argue that the
long-range nature of the dipolar forces plays a key role in yielding
ferroelectric order, and that this is not explicitly included in the simple
interpretation given above. More importantly, Zhang and Widom propose that,
despite the strong frustration present in randomly frozen systems, long-range
ferroelectric order is possible above a critical density. Their work implies
that the spin-glass behavior observed in ferrofluids \cite{luo} results from
the low concentration of magnetic particles, whereas the ferroelectric liquid
crystalline phase found in computer simulations of dipolar fluids
\cite{wei,weis} arises because of the high particle density considered. In the
present letter we examine the validity of this argument and address the
general question {\em ``Can long-range ferroelectric order spontaneously
arise in a system without fined-tuned positional correlations?''}

We investigate the behavior of dense spatially disordered dipolar
systems using constant temperature molecular dynamics (MD) and Monte
Carlo (MC) simulations. Systems where the dipole vector has one, two and three
components are considered. The first two of these are commonly referred to as
the Ising and XY models and for notational simplicity we shall refer to the
three component dipole as the XYZ model. In all cases the pair potential,
$u(12)$, is of the generic form
$$u(12) = 4\varepsilon(\sigma/r)^{12} + u_{DD}(12)\ ,$$
where $4\varepsilon(\sigma/r)^{12}$ and
$$u_{DD}(12) = -3(\bbox{\mu}_{1} \cdot {\bf r} )(\bbox{\mu}_{2} \cdot
{\bf r})/r^{5}
+ \bbox{\mu}_{1} \cdot \bbox{\mu}_{2}/r^{3}$$
\noindent
are the soft-sphere and dipole-dipole interactions. The parameters
$\varepsilon$ and $\sigma$ are the fundamental units of energy and length,
$\bbox{\mu}_{i}$ is the dipole of particle $i$ and ${\bf r}$ is the
intermolecular vector. The long-range
dipolar interactions were taken into account using periodic boundary
conditions  and the Ewald summation method assuming a perfectly conducting
surrounding continuum \cite{wei,deleeuw,kusalik,cont}. The existence of a
ferroelectric phase can be detected by calculating the average polarization
$P$ per particle defined as
$P =    1/N \langle | \sum_{i=1}^{N}
\hat{\bbox{\mu}}_{i}\cdot\hat{{\bf d}}| \rangle$,
where $\hat{{\bf d}}$ is a unit vector in the direction of the total
instantaneous moment, ${\bf M}= \sum_{i=1}^{N}
\bbox{\mu}_{i}$, and $N$ is the number of particles in the system.

It is convenient to characterize dipolar soft-sphere systems by the reduced
density, $\rho^{*} = N \sigma^{3}/V$, the reduced temperature, $T^{*} = k_{\rm
B}T/\varepsilon$, and the reduced dipole moment, $\mu^{*} =
(\mu^{2}/\varepsilon \sigma^{3})^{1/2}$, where $V$ is the volume of the
sample, $T$ is the absolute temperature and $k_{\rm B}$ is the Boltzmann
constant. All results explicitly presented are for $\mu^{*} = 4$ and $\rho^{*}
= 0.8$. This density is well within the range for which Zhang and Widom
predict a ferroelectric phase \cite{crit}. For $\mu^{*} = 4$ and $\rho^{*} =
0.8$, Zhang and Widom predict a ferroelectric phase for the Ising case if
$T^{*} \leq 35.2$ and for the XYZ model if $T^{*} \leq 4.8$.

We first consider frozen systems. Suitable spatially disordered configurations
were generated by carrying out an MD simulation of a soft-sphere fluid at
$T^{*}=10.5$ and $\rho^{*}=0.8$. Fluid-like configurations were then selected
at random for dipolar rotational MD simulations. Following this approach we
could obtain a frozen state at a much higher density than is possible from a
random parking algorithm. Unfortunately, it is impossible to have a truly
``random" and structureless spatial configuration (i.e., with the radial
distribution function $g(r)$ equal to one for $r \ge \sigma$ \cite{widom}) at
this density. However, at $T^{*}=10.5$ the local structure in the soft-sphere
fluid is weak and very short-ranged.

Polarization results for randomly frozen systems are shown in Fig. 1
\cite{fig1}. The XY and XYZ values were obtained with MD simulations. The
discrete nature of the Ising model renders it inappropriate for MD so MC
calculations were used. The average polarization obtained at the lowest
temperature where equilibrium could be achieved, $T^{*}_{\rm min}$, is plotted
vs $1/N$. The values of $T^{*}_{\rm min}$ are 10.0, 4.0, and 3.5 for the
Ising, XY and XYZ models, respectively. Below these temperatures MD or MC runs
for the same configuration started from perfectly ordered and disordered
states (i.e., two replicas) did not converge to the same result within a
reasonable computation time (i.e., about a week). Possibly with greater
computational effort $T^{*}_{\rm min}$ could be pushed a little lower, but the
values given above are well within the range where Zhang and Widom predict a
ferroelectric phase. For the Ising model the equilibrated system at
$T^{*}=10.0$ is nearly 100\% polarized. Moreover, for the Ising case the
polarization at $T^{*}_{\rm min}$ exhibits very little number dependence and
certainly does not appear to vanish in the thermodynamic limit. This, together
with the plot of $P$ vs $T^{*}$ and heat capacity calculations (see Fig. 3
below), strongly suggests that ferroelectric order develops spontaneously in
the spatially disordered Ising system with the transition occurring at
$T^{*}\approx 25$ for $\rho^*=0.8$. We note that this transition temperature
is much lower than the value ($T^{*}=35.2$) predicted by Zhang and Widom.

The situation for the XY and XYZ models is very different. Although
significant polarization was observed at finite temperatures, $P$ decreases
monotonously with increasing system size and appears to approach zero for an
infinite system. Furthermore, the behavior of various spin-glass correlation
functions and susceptibilities \cite{BY} suggests that both systems are
entering an anisotropic spin-glass phase at nonzero temperature
\cite{gingras}, and that the observed polarization for the XY and XYZ models
is due to a combination of short-range ferroelectric correlations and
finite-size effects. Test calculations for the XYZ model using a denser frozen
soft-sphere configuration (i.e., $\rho^{*}=1.05$ \cite{crit}, $T^{*}=10.5$)
also showed no long-range ferroelectric order. In brief, for the randomly
frozen XY and XYZ models we find no evidence of a ferroelectric state in the
thermodynamic limit. This clearly disagrees with the calculations of Zhang and
Widom which for the XYZ model predict a stable ferroelectric phase well within
the temperature-density range considered here.

In order to gain further insight into the nature of ferroelectric order (or
the lack thereof) in spatially random systems, we consider a ``toy model''
where the translational motion is completely decoupled from the dipolar
interactions. The soft-sphere fluid acts as a ``substrate'' which moves at a
{\em fixed translational} temperature independent of the embedded dipoles. The
dipoles themselves interact and are equilibrated at a different {\em
rotational} temperature. Of course, the ``equilibrium'' state achieved by the
dipoles will depend on the underlying motion of the substrate. This model is
similar in spirit to those used in recent studies of non-equilibrium phase
transitions in magnetic systems subject to Levy flights \cite{levy}. It must
be emphasized that this technique is presented only as a useful simulation
tool and we do not imply any real physical mechanism for the decoupling. The
moving substrate is a means of simulating dipolar systems in a {\em
dynamically} random medium that lacks any specific spatial correlations which
may favor ferroelectric ordering. The translational diffusion rate of the
substrate can be controlled by the particle mass. Extrapolation to infinite
mass should provide information about the randomly frozen system.

In Fig. 2, we have plotted $P$ vs $T^{*}$ (rotational) for the XYZ model.
Here, the decoupled substrate is a soft-sphere fluid again at $\rho^*=0.8$ and
$T^{*} ({\rm translational}) = 10.5$. It is convenient to introduce the
reduced mass $m^{*}=m/m^{\prime}$, where $m^{\prime}$ is a reference mass
defined such that the reduced simulation timestep $\Delta t^{*} \equiv
(\epsilon/m^{\prime}\sigma^{2})^{1/2} \Delta t = 0.00125$. Figure 2 includes
results for $m^{*}=1$, 5 and 10. Spontaneous polarization
develops for all systems and the temperature at which $P$ begins to grow
decreases with increasing mass. For example, from the $P$ vs $T^{*}$ plot
there appears to be a ferroelectric transition at $T^{*} \approx 2$ for
$m^{*}=1$. To verify that this is a real thermodynamic transition, we have
calculated the Binder ratio \cite{BY}, $g_{\rm Binder} \equiv
(5/2)-(3/2)\langle|{\bf M}|^4\rangle/\langle|{\bf M}|^2\rangle^2$, for systems
with 64, 108 and 256 particles. A plot of $g_{\rm Binder}$ vs $T^{*}$ (see
Fig. 2, inset) shows a clear crossing, and hence a transition, at $T^{*}=1.9$.
Simulations were also carried out with $m^{*}=20$ but no significant
orientational order was found above $T^{*}=0.1$. Very slow convergence
prevented calculations at lower temperatures.

We have also investigated dynamically disordered XY and Ising systems. The XY
model behaves much like the XYZ system described above. For the Ising case,
rotational dynamics cannot be used and a suitable Monte Carlo scheme which
allowed the substrate to move independent of the Ising dipoles was employed.
$P$ vs $T^{*}$ results for $m^{*}=1$, $m^{*}=5$, and the
randomly frozen system ($m^{*}=\infty$) are shown in Fig. 3. We see
that for the Ising model the ordering behavior is essentially independent of
mass and that the results for the dynamically disordered system with $m^{*}=5$
lie very close the those for the randomly frozen case. Heat capacities
obtained by differentiating the average dipolar energy with respect to the
rotational temperature are also shown in Fig. 3. The
randomly frozen and $m^{*}=5$ results are very similar and indicate a phase
transition at $T^{*} \approx 25$.

The dependence of the ferroelectric transition temperature on particle mass is
shown in Fig. 4. The transition temperatures were estimated from the heat
capacities and results are included for the XY and XYZ
models. Results for the Ising system are not plotted because the transition
temperature is essentially independent of the mass. We see that as the mass
increases the transition temperature drops for both the XY and XYZ models. As
noted earlier, for large masses and low temperatures convergence becomes
prohibitively slow, but it seems reasonable to assume that the graph would
simply continue with the transition temperature approaching zero in the
infinite mass limit. This is consistent with the fact that we did not find a
ferroelectric phase for randomly frozen systems at finite temperatures.

The present and previous results can be interpreted as follows. It is useful
to divide the local field, ${\bf E}_{\rm local}$, experienced by a particle in
an infinite medium into two parts such that, ${\bf E}_{\rm local} = {\bf R} +
{\bf E}$, where ${\bf R}$ is a reaction field contribution \cite{wpp,deleeuw}
and ${\bf
E}$ is everything else. The reaction field arises from the long-range nature
of the dipolar interactions; it is {\em independent} of the spatial
correlations and favors ferroelectric order. The remaining contribution,
${\bf E}$, is sensitive to positional correlations and may or may not favor
ferroelectric order. Thus if ${\bf R}$ dominates a ferroelectric phase is to
be expected, but if ${\bf E}$ is important the existence or nonexistence of a
ferroelectric phase will depend on the details of the spatial correlations.
This simple picture allows us to rationalize the various systems considered.
For fully coupled dipolar fluids \cite{wei,weis} the short-range spatial
correlations (and hence ${\bf E}$) can adjust (i.e., become tetragonal-I-like)
in order to favor (or at least allow) a ferroelectric state. On the other
hand, for randomly frozen systems the spatial correlations cannot adjust and
at equilibrium ${\bf E}$ dominates ${\bf R}$ frustrating the formation of a
ferroelectric phase except for the Ising case. Apparently, the discrete nature
of the Ising model makes it much less susceptible to the development of
disordering fields than are the XY and XYZ systems. The reaction field
dominates in the Ising system and gives rise to a ferroelectric state.

The behavior of the dynamically disordered systems can also be understood. If
the translational diffusion of the substrate is sufficiently fast compared to
the dipolar reorientational time, the extent of spatially dependent
frustrating correlations is limited, ${\bf R}$ can prevail over ${\bf E}$, and
a ferroelectric phase is possible. As the mass of a substrate particle is
decreased, the translational motion becomes faster and the above condition is
met at higher and higher rotational temperatures. Thus the observed transition
temperatures increase with decreasing mass. As the substrate particles become
sufficiently light, the transition temperature is determined only by the
reaction field and hence becomes independent of mass. In fact, $m^{*}=1$ gives
essentially this limiting behavior and reducing the mass further has little
effect on the transition temperature.

In conclusion, the answer to the question raised at the outset is a subtle
one. Our results for the frozen Ising system and for the dynamically
disordered XY and XYZ models clearly demonstrate that it is possible to have
ferroelectric states without fine-tuned positional correlations. However, they
also show that if a ferroelectric phase is to exist in a positionally random
system, the long-range spatially-independent correlations arising through the
reaction field must dominate the shorter-ranged position-sensitive
correlations which generally act to frustrate ferroelectric order. For the
Ising system the discrete nature of the interaction (i.e., there is just not
much opportunity for favorable interactions among neighbors) limits the build
up of short-range disordering correlations and there is a clear ferroelectric
transition. For the dynamically random systems the orientational correlations
opposing ferroelectric order are reduced by the motion of the substrate
resulting in a ferroelectric phase. On the other hand, for the randomly frozen
XY and XYZ models we find no indication of a ferroelectric phase at finite
temperature. Rather, the disordering fields dominate and these systems appear
to form nonferroelectric spin glasses at low temperature. Evidence for this is
provided both by our direct simulations of frozen systems and by the
extrapolation of the dynamically disordered results to infinite mass. This
conclusion must remain a little tentative because direct MD simulations at
very low temperatures are not practical. Nevertheless, a ferroelectric phase
in the randomly frozen XY and XYZ models seems highly unlikely.

\acknowledgments

We thank Z. R\'acz, M. Widom and  H. Zhang for useful discussions. The
financial support of the National Science and Engineering Research Council of
Canada is gratefully acknowledged.

\begin{figure}
\caption{
The polarization $P$ at $T^{*}_{\rm min}$ vs $100/N$ for the randomly
frozen Ising (circles), XY (squares) and XYZ (triangles). Results are
included for 108 (XYZ only), 256, 392 (XYZ only), 500 and 864 particles.}
\end{figure}

\begin{figure}
\caption{
$P$ vs $T^{*}$ (rotational) for dynamically random $XYZ$
systems. The squares, triangles and circles are for $m^{*}=1$, 5 and 10,
respectively. The error bars represent one estimated standard deviation.
$g_{\rm Binder}$ vs $T^{*}$ (rotational) is shown in the inset for $N=64$
(squares), 108 (triangles) and 256 (circles) particles.}
\end{figure}

\begin{figure}
\caption{
$P$ vs $T^{*}$ (rotational) for the Ising model. Results are shown for
dynamically random systems with $m^{*}=1$ (squares) and $m^{*}=5$ (triangles)
and for the randomly frozen case (circles). The heat capacities per particle,
$C_{v}/N$, are plotted vs $T^{*}$ (rotational) in the inset.}
\end{figure}

\begin{figure}
\caption{
The mass dependence of the disordered to ferroelectric transition temperature
$T^{*}_{F}$ (rotational). The squares and triangles are results for the XY and
XYZ models, respectively. The values of $T^{*}_{F}$ were obtained from plots
of the heat capacity, $C_{v}/N$, vs $T^{*}$ (rotational) and a typical example
(XYZ model, $m^{*}=1$) is shown in the inset. The error bars represent
estimated uncertainties in the peak position.}
\end{figure}


\begin{references}
\bibitem{cont} For the present dense strongly dipolar systems the dielectric
constants are sufficiently large that the reaction field is essentially
independent of the exact value \cite{deleeuw}. Therefore, conducting boundary
conditions are an appropriate choice.
\bibitem{fig1} The results given in Fig. 1 are for typical frozen
configurations. Tests show that the actual values of $P$ vary a little from
configuration to configuration but that the same general trend is found.
Ideally, accurate values of $P$ could be obtained by averaging over many
frozen configurations for each $N$. However, this would be very expensive
computationally and is not necessary here; we are only interested in the
limiting value of $P$ as $N$ becomes large and not in its detailed $N$
dependence.
\end{references}
\end{document}